\documentclass[a4paper,11pt]{article}
\usepackage{bm,mathrsfs,amsmath,amsthm,amssymb,eepic,amscd,longtable,array,graphicx}
\usepackage[dvips]{color}
\newcommand{\nc}{\newcommand}
\nc{\rnc}{\renewcommand}
\nc{\nn}{\nonumber}
\nc{\del}{{\partial}}
\rnc{\Im}{{\rm{Im}\,}}
\rnc{\Re}{{\rm{Re}\,}}
\nc{\db}{\displaybreak[0]\\}
\nc{\bra}{\langle}
\nc{\ket}{\rangle}

\nc{\lam}{\lambda}
\nc{\g}{{\mathfrak{g}}}
\nc{\zb}{\bar{z}}
\nc{\hb}{\bar{h}}
\nc{\J}{\mathcal{J}}
\nc{\su}{\widehat{\mathfrak{su}}(2)_k}

\nc{\tcr}{\textcolor{red}}

\numberwithin{equation}{section}
\numberwithin{lemma}{section}
\numberwithin{proposition}{section}
\numberwithin{theorem}{section}
\numberwithin{corollary}{section}
\numberwithin{conjecture}{section}

\begin{document}%
%%%%%%%%%%%%%%%%%%%%%%%%%%%%%%%%%%%%%%%%%%%%%%%%%%%%%%%%%
%TITLE
%%%%%%%%%%%%%%%%%%%%%%%%%%%%%%%%%%%%%%%%%%%%%%%%%%%%%%%%%
%
\title{Multiple Schramm-Loewner evolutions for coset Wess-Zumino-Witten models}

\author{
\\
Yoshiki Fukusumi$^1$\thanks{E-mail: y.fukusumi@issp.u-tokyo.ac.jp} 
\\\\
{\it The Institute for Solid State Physics, University of Tokyo}\\
{\it Kashiwa, Chiba 277-8581, Japan} \\
}
%\date{}
%%%%%%TEXT START%%%%%% 
\maketitle
%%%%%%%%%%%%%%%%%%%%%%%%%%%%%%%%%%%%%%%%%%%%%%%%%%%%%%%%%
%

%
\begin{abstract}
We formulate multiple Schramm-Loewner evolutions (SLEs) for 
coset Wess-Zumino-Witten (WZW) models. The resultant SLEs
may describe  the critical behavior of multiple interfaces 
for the 2D statistical mechanics models whose critical 
properties are classified by coset WZW models. The SLEs
are essentially characterized by multiple Brownian motions 
on a Lie group manifold as well as those on the real axis.
The drift terms of the Brownian motions, which come from interactions of interfaces,
are explicitly determined by imposing a martingale condition
on correlation functions among 
boundary condition changing operators.
As a concrete example, we formulate multiple SLE on the
$Z(n)$ parafermion model and calculate the crossing probability which
is closely related to 3-SLE drift terms.
\end{abstract}
%
%%%%%%%%%%%%%%%%%%%%%%%%%%%%%%%%%%%%%%%%%%%%%%%%%%
\section{Introduction}
%%%%%%%%%%%%%%%%%%%%%%%%%%%%%%%%%%%%%%%%%%%%%%%%%%
%
Schramm-Loewner evolution (SLE) is a growth process
describing  critical properties of interfaces for scaling 
limits of 2D lattice systems at criticality \cite{S1} 
(see for instance \cite{Smi, GC} as a realization of SLE
traces in some 2D lattice systems). SLE is described by 
a differential equation with a 1D Brownian motion as 
a driving force.  The connection between SLE and conformal
field theory (CFT) is well established by imposing a martingale
condition on the CFT correlation functions \cite{BB}
(see also \cite{BB0}). 

There exist several versions of SLEs.  The first one 
is obtained by changing the geometry of the domain: the chordal, 
radial and dipolar SLE are defined on the upper half plane, 
the unit disk and a strip, respectively \cite{S1,BBH}. These SLEs
give a single random curve starting from the origin to infinity for
the chordal case,
from the origin on the lower boundary to a random point on the upper boudary 
for the dipolar case, and from a point on the boundary of the unit disk
to the center of the disk for the radial case.

The second one is the multiple version of the SLE describing
multiple random  interfaces \cite{BBK} (see \cite{Dub} for 
another description). In this case, the driving forces are 
multiple Brownian motions on the real axis, which possess
drift terms reflecting the effect of interactions among
the SLE traces. The drift terms can be explicitly 
given by CFT correlation functions among the boundary
condition changing (bcc) operators.

The third one, which is mainly discussed in this paper, is SLEs 
with additional internal degrees of freedom such as spins and colors. 
This can be accomplished by considering an additional Brownian 
motion with a Lie group manifold so that 
the resultant SLE is compatible with Wess-Zumino-Witten (WZW) models 
\cite{BGLW,ABI}. An extension to the multiple version of this
type of the SLE has also been achieved in \cite{Sak}.

In all of these generalization of SLE, the martingale condition clarifies the correspondence between
SLE and CFT. It ensures the welldefined construction of boundary conformal field theory (BCFT) correlation function
as distribution function of SLE as we will briefly revisit in the next section.\footnote{YF thanks Raoul Santachiara
pointing out that notion.}

The present paper is devoted to an extension of the single SLE
compatible with coset WZW models \cite{Naz1,Naz2} to the multiple 
version. There exist varaeties of 2D critical statistical models
corresponding to coset WZW models \cite{JMO,DSZ,Fen}. One of the most fundamental models
is the $Z(n)$ spin model corresponding to the $SU(2)_n/U(1)_n$-WZW model \cite{ZF}.
As a concrete application, considering the triple SLEs and 
solving the Knizhnik-Zamolodchikov (KZ) equation \cite{KZ} satisfied by
the correlation functions among bcc operators,
we formulate the crossing probabilities of spin clusters of the
$Z(n)$ spin models.

This paper is orgnized as follows. In the next section, we give more detail of the
general framework of the general construction of multiple SLE on the coset WZW models. 
An application 
to the $Z(n)$ spin models of our framework is devoted in Section 3. Some technical details
are deferred to the appendices. An interesting relation for the crossing 
probability between  
the $SU(2)_2$-WZW model and the Ising model is also discussed in the
appendix C.

%%%

%%%%%%%%%%%%%%%%%%%%%%%%%%%%%%%%%%%%%%%%%
\subsection{Summary}
%%%%%%%%%%%%%%%%%%%%%%%%%%%%%%%%%%%%%%%%%
%
What we have shown in this paper is the possibility of  multiple SLE
on $G/A$ coset WZW model under the existence of 1-SLE.
Our discussion mainly relies on three assumptions related to BCFT. 
They are the existence of primary fields, and the form of OPE, and
the stochastic derivative of the primary operators.
Under these assumptions, we calculated the stochastic derivative of the
correlation functions (in that sense, what we need is the calculation of
the correlation functions. Therefore these assumptions are sufficient conditions).
First, we assume the form of the multiple SLE on the complex plane and
the $G/A$ group manifold \cite{Sak, Naz1},
\begin{alignat}{2}
&dg_t(z)=\sum_{\alpha=1}^m \frac{2 dq_\alpha}{g_t(z)-x_{\alpha t}},&\qquad&
d x_{\alpha t}=\sqrt{\kappa}d\xi_{\alpha t}+dF_{\alpha t}  \nn \\
& d\theta_t^a(z)=\sum_{\alpha=1}^m\frac{dp^a_{\alpha t}}{z-x_{\alpha t}},
&& dp_{\alpha t}^a=\sqrt{\tau} d\vartheta_{\alpha t}^a+dG_{\alpha t}^a 
\quad (1\le a\le\dim \g).
\label{sle-def}
\end{alignat}
The expectation value and variance of Brownian motion $\xi_{\alpha t}, \vartheta^a_{\alpha t}$ are, respectively given by 
$\mathbf{E}[d\xi_{\alpha t} d\xi_{\beta t}]=\delta_{\alpha\beta}dq_\alpha,\mathbf{E}[d\vartheta^a_{\alpha t} d\vartheta_{\beta t}^b]=
K[t^a, t^b]\delta_{\alpha\beta}dq_\alpha$, where $K[ , ]$ is killing form. 

Then we assign the martingale condition to arbitrary correlation functions 
described by the product of the primary fields on the upper half plane.
It is summarized to \eqref{ns-condition}.
Applying all assumptions, the explicit forms of the drift terms are derived,
\begin{align}
&dF_{\alpha t} =\kappa dq_\alpha \del_{x_{\alpha t}} \log Z_t+
2\sum_{\substack{\beta=1\\\beta\ne \alpha}}^m
\frac{dq_\beta }{x_{\alpha t}-x_{\beta t}}, \quad
dG_{\alpha t}^a =\frac{\tau}{Z_t}\sum_{\substack{\beta=1\\\beta\ne\alpha}}^m
\frac{t_{\lam_\beta}^a Z_t}{x_{\beta t}-x_{\alpha t}}dq_\alpha,
\label{driving-sum}
\end{align}
where $Z_t$ is a partition function costructed by the bcc operators:
\begin{equation}
Z_t=
\bra\psi_{\Lambda}(x_{1t})\cdots\psi_{\Lambda}(x_{mt}) 
\psi_{\lam_{m+1}}(\infty)\ket.
\label{CFT-cor}
\end{equation}
$\psi_{\lam_{m+1}}(\infty)$ is determined by fusion rule and
the analysis of 2-SLE as we will see in the next section.
The basic idea is random time changing and the analysis
of the Bessel process.
After determination of $\psi_{\lam_{m+1}}(\infty)$ corresponding
to the actual situations, it may contain several conformal blocks.
We will introduce some idea to extract "physical" process
based on the correspondence of 2-SLE and the fusion rule of BCFT.
The key calculation tool is the bosonization or Dotsenko-Fateev (DF) integral.

For last but not least remark, what we assumed is the existence of 1-SLE of coset WZW models.
In other words, we showed the validity of generalization of 1-SLE to multiple SLE.
The strength of our formulation is the generality of the construction of probability measure of process which is
consistent with BCFT.

%%%
%%%%%%%%%%%%%%%%%%%%%%%%%%%%%%%%%%%%%%%%%%%%%%%%%
\section{General consideration of the multiple SLEs on the coset WZW models}
%%%%%%%%%%%%%%%%%%%%%%%%%%%%%%%%%%%%%%%%%%%%%%%%%%
%
In this section, we briefly show general framework
for the construction of multiple SLEs on the WZW models. 
%
%
%%%%%%%%%%%%%%%%%%%%%%%%%%%%%%%%%%%%%%%%%%%%%%%%%%
\subsection{Multiple SLEs with Lie algebraic symmetries}
%%%%%%%%%%%%%%%%%%%%%%%%%%%%%%%%%%%%%%%%%%%%%%%%%%

First, we assume the form of multiple $m$-SLE on $G/A$ Lie group manifold and upper half plane $H=\{ z\in \mathbf{C}| \Im z >0 \}$.
For simplicity, we assume $A$ is generated by Cartan subalgebra of $G$.  
The multiple SLE is a stochastic conformal map $g_{t}:H\rightarrow H\setminus K_t$
with Brownian motion on $G/A$ Lie group manifold represented by $\exp (\sum_{a}^{\dim \g-\dim\mathfrak{a}}{s^a \theta_{t}^{a} (z)})$ \cite{Naz1}, 
\begin{alignat}{2}
&dg_t(z)=\sum_{\alpha=1}^m \frac{2 dq_\alpha}{g_t(z)-x_{\alpha t}},&\qquad&
d x_{\alpha t}=\sqrt{\kappa}d\xi_{\alpha t}+dF_{\alpha t}  \quad(1\le\alpha\le m) \nn \\
& d\theta_t^a(z)=\sum_{\alpha=1}^m\frac{dp^a_{\alpha t}}{g_t(z)-x_{\alpha t}},
&& dp_{\alpha t}^a=\sqrt{\tau} d\vartheta_{\alpha t}^a+dG_{\alpha t}^a 
\quad (1\le a\le\dim \g-\dim \mathfrak{a}).
\label{sle-sum}
\end{alignat}

The expectation value and variance of Brownian motion $\xi_{\alpha t}, \vartheta^a_{\alpha t}$ are given by 
$\mathbf{E}[d\xi_{\alpha t} d\xi_{\beta t}]=\delta_{\alpha\beta}dq_\alpha,\mathbf{E}[d\vartheta^a_{\alpha t} d\vartheta_{\beta t}^b]=
\delta_{ab}\delta_{\alpha\beta}dq_\alpha.$
Moreover, we assume the map satisfying
$g_t (\gamma_{\alpha t})=x_{\alpha t}$.
Usually, we think about $H\setminus K_t$ as a fractal object generated by $\gamma_{\alpha t}$. 

At this stage, we can choose arbitrary value of $\kappa, \tau$ and drift terms $dF_{\alpha t}, dG_{\alpha t}^a$.
To determine them, another assumption is necessary for the SLE/BCFT correspondence. 
In other words, it is needed to fix these parameters for the description of the specific models.
For this purpose, we assume the existence of the primary fields and their stochastic
derivatives on $H$ (if necessary, by using duplication trick):
\begin{align}
d \phi(y)&=\frac{\del \phi(y)}{\del y}\sum_{\alpha=1}^m\frac{\del y}{\del q_\alpha} 
dq_\alpha+\sum_{a=1}^{\dim \g-\dim\mathfrak{a}} s^a d\theta^a(y)
              \phi(y)
+\frac{1}{2}\sum_{a=1}^{\dim \g-\dim\mathfrak{a}} s^a d\theta^a(y) 
                   \sum_{b=1}^{\dim \g-\dim\mathfrak{a}} s^b d\theta^b(y)
   \phi(y)                      \nn \\
&=\sum_{\alpha=1}^m\left[ \frac{2dq_\alpha\del_y}{y-x_\alpha}+
     \sum_{a=1}^{\dim \g-\dim\mathfrak{a}}
     \frac{dp_\alpha^a s^a}{y-x_\alpha}+
    \frac{\tau}{2}\sum_{a=1}^{\dim \g-\dim\mathfrak{a}}
     \frac{dq_\alpha s^as^a}{(y-x_\alpha)^2}
\right]\phi(y),
\end{align}
where $\phi(y)$ is an arbitrary primary field.
Moreover, by the assumption that SLE curves are represented by bcc operator $\psi$, 
the stochastic derivative of them are given by,
\begin{align}
d\psi_{\alpha}(x_\alpha)&=
\left[dq_\alpha\frac{\kappa}{2}\del_{x_\alpha}^2+dx_\alpha\del_{x_\alpha}  
+\sum_{a=1}^{\dim \g-\dim\mathfrak{a}} s^a d\theta^a(x_\alpha)
+\frac{1}{2}\sum_{a=1}^{\dim \g-\dim\mathfrak{a}} s^a d\theta^a(x_\alpha) 
                   \sum_{b=1}^{\dim \g-\dim\mathfrak{a}} s^b d\theta^b(x_\alpha)
\right]
   \psi_{\alpha}(x_\alpha)           
\nn \\
&=\left[dq_\alpha\frac{\kappa}{2}\del_{x_\alpha}^2+
d x_\alpha \del_{x_\alpha}
+ \sum_{a=1}^{\dim \g-\dim\mathfrak{a}}\sum_{\substack{\beta=1\\ \beta\ne\alpha}}^m
 \left(\frac{dp_\beta^a s_{\alpha}^a}{x_\alpha-x_\beta}+
    \frac{\tau}{2}
     \frac{dq_\beta s_\alpha^as_\alpha^a}{(x_\alpha-x_\beta)^2}\right)
\right]\psi_{\alpha}(x_\alpha).
\end{align}
Then we assign martingale condition to the following arbitrary operators,
\begin{equation}
\mathcal{M}_t=
\prod_{j=1}^{2n}\left( \frac{\del y_{jt}}{\del z_j} \right)^{h_{j}}
\frac{\bra\prod_{j=1}^{2n} \phi_{j}(y_{jt}) \psi_{\Lambda}(x_{1t})\cdots
\psi_{\Lambda}(x_{mt}) \psi_{\lam_{m+1}}(\infty)\ket}
{\bra\psi_{\Lambda}(x_{1t})\cdots\psi_{\Lambda}(x_{mt}) \psi_{\lam_{m+1}}(\infty)\ket }
=\frac{J^{\phi}_{t}Z^{\phi}_{t}}{Z_{t}}.
\label{observable}
\end{equation}
The stochastic Ito caluculus leads to,
\begin{equation}
d\mathcal{M}_t=d\left(\frac{J_t^\phi Z_t^\phi}{Z_t}\right)
=\frac{d(J_t^\phi Z_t^\phi)}{Z_t}-\frac{J_t^\phi Z_t^\phi dZ_t}{Z_t^2}
-\frac{d(J_t^\phi Z_t^\phi )dZ_t}{Z_t^2}+
\frac{J_t^\phi Z_t^\phi (dZ_t)^2}{Z_t^3}.
\label{mart-SLE}
\end{equation}
Each terms can be calculated by substitution of the following forms (for more detailed calculation,
see \cite{Sak}),
\begin{align}
dZ_t=&
\sum_{\alpha=1}^m dq_\alpha\left(
\frac{\kappa}{2}\widetilde{\mathcal{L}}_{\alpha,-1}^2-2\widetilde{\mathcal{L}}_{\alpha,-2}+
\frac{\tau}{2}\sum_{a=1}^{\dim \g}\widetilde{\J}_{\alpha,-1}^a
 \widetilde{\J}_{\alpha,-1}^a-\frac{\tau}{2}\sum_{b=1}^{\dim \mathfrak{a}}\widetilde{\J}_{\mathfrak{a},\alpha,-1}^b
 \widetilde{\J}_{\mathfrak{a},\alpha,-1}^a
\right)  Z_t\nn \\
&
+\sum_{\alpha=1}^m \left(
2\widetilde{\mathcal{L}}_{\alpha,-2} dq_\alpha
+\widetilde{\mathcal{L}}_{\alpha,-1} dx_\alpha
+\sum_{a=1}^{\dim \g-\dim\mathfrak{a}}\widetilde{\J}_{\alpha,-1}^a dp_\alpha^a \right) Z_t,
\label{purepartion}
\end{align}
\begin{align}
\frac{d(J_t^\phi Z_t^\phi)}{J_t^\phi}=&
\sum_{\alpha=1}^m dq_\alpha\left(
\frac{\kappa}{2}\mathcal{L}_{\alpha,-1}^2-2\mathcal{L}_{\alpha,-2}+
\frac{\tau}{2}\sum_{a=1}^{\dim \g-\dim\mathfrak{a}}\J_{\alpha,-1}^a\J_{\alpha,-1}^a
\right)  Z_t^\phi\nn \\
&
+\sum_{\alpha=1}^m \left(
2\widetilde{\mathcal{L}}_{\alpha,-2} dq_\alpha
+\mathcal{L}_{\alpha,-1} dx_\alpha
+\sum_{a=1}^{\dim \g-\dim\mathfrak{a}}\J_{\alpha,-1}^a dp_\alpha^a \right) Z_t^\phi
\nn \\=&
\sum_{\alpha=1}^m dq_\alpha\left(
\frac{\kappa}{2}\mathcal{L}_{\alpha,-1}^2-2\mathcal{L}_{\alpha,-2}+
\frac{\tau}{2}\sum_{a=1}^{\dim \g}\J_{\alpha,-1}^a\J_{\alpha,-1}^a
-\frac{\tau}{2}\sum_{a=1}^{\dim \mathfrak{a}}\J_{\alpha,-1}^a\J_{\alpha,-1}^a
\right)  Z_t^\phi\nn \\
&
+\sum_{\alpha=1}^m \left(
2\widetilde{\mathcal{L}}_{\alpha,-2} dq_\alpha
+\mathcal{L}_{\alpha,-1} dx_\alpha
+\sum_{a=1}^{\dim \g-\dim\mathfrak{a}}\J_{\alpha,-1}^a dp_\alpha^a \right) Z_t^\phi
\label{martingale}
\end{align}
with,
\begin{align}
&\mathcal{L}_{\alpha,-l}=\sum_{j=1}^n 
\left(
\frac{(l-1)h_j}{(y_j-x_\alpha)^l}-\frac{\del_{y_j}}{(y_j-x_\alpha)^{l-1}}
\right)+
\sum_{\substack{\beta=1\\\beta\ne\alpha}}^m
\left(
\frac{(l-1)h_\beta}{(x_\beta-x_\alpha)^l}-\frac{\del_{x_\beta}}{(x_\beta-x_\alpha)^{l-1}}
\right), \nn \\
&\widetilde{\mathcal{L}}_{\alpha,-l}=\sum_{\substack{\beta=1\\\beta\ne\alpha}}^m
\left(
\frac{(l-1)h_\beta}{(x_\beta-x_\alpha)^l}-\frac{\del_{x_\beta}}{(x_\beta-x_\alpha)^{l-1}}
\right), \nn \\
&\J_{\alpha,-l}^{a}=\sum_{j=1}^n\frac{s_{\lam_j}^a}{(y_j-x_\alpha)^l}
+\sum_{\substack{\beta=1\\\beta\ne\alpha}}^m
\frac{s_{\lam_\beta}^a}{(x_\beta-x_\alpha)^l}, \quad (1\le a\le\dim \g), \nn \\
&\widetilde{\J}_{\alpha,-l}^{a}=\sum_{\substack{\beta=1\\\beta\ne\alpha}}^m
\frac{s_{\lam_\beta}^a}{(x_\beta-x_\alpha)^l},\quad (1\le a\le\dim \g),
\label{LJ-op2}
\end{align}
For 1-SLE, only the first term of \eqref{observable} survives. We discuss the BCFT meaning of that equation
in the next subsection.
If we suppose 1-SLE constructed, then we get the maritingale
condition as the constraint of the drift terms,
\begin{align}
\mathbf{E}[d\mathcal{M}_t]=&J_t^\phi \sum_{\alpha=1}^m\left(dF_\alpha -
\kappa dq_\alpha  \del_{x_\alpha}\log Z_t
-2\sum_{\substack{\beta=1\\\beta\ne \alpha}}^m
\frac{dq_\beta }{x_\alpha-x_\beta}
\right)\del_{x_\alpha}\left(\frac{Z_t^\phi}{Z_t}\right)  \nn \\
&\qquad+
\frac{J_t^\phi}{Z_t}
\sum_{\alpha=1}^m\sum_{a=1}^{\dim \g -\dim\mathfrak{a}}
\left[
\left(dG_\alpha^a-
\frac{\tau dq_\alpha}{Z_t}\widetilde{J}^{a}_{\alpha,-1}Z_t
\right)\left(J_{\alpha,-1}^a Z_t^\phi-
\frac{Z_t^\phi\widetilde{\J}_{\alpha,-1}^a}{Z_t}Z_t\right)
\right],
\end{align}
We discuss the meaning of the pure partition function in the
following subsection.

%%%%%%%%%%%%%%%%%%%%%%%%%%%%%% 
\subsection{The importance of constructing 1-SLE}
%%%%%%%%%%%%%%%%%%%%%%%%%%%%%%%
For 1-SLE case, the martingale conditions are,
\begin{equation}
\left(
\frac{\kappa}{2}\mathcal{L}_{\alpha,-1}^2-2 \mathcal{L}_{\alpha,-2}
+\frac{\tau}{2}
\sum_{a=1}^{\dim \g}\J_{\alpha,-1}^{a}\J_{\alpha,-1}^{a}-\frac{\tau}{2}
\sum_{a=1}^{\dim \mathfrak{a}}\J_{\alpha,-1}^{a}\J_{\alpha,-1}^{a}\right)Z_t^\phi=0 
\quad (\alpha=1),
\label{1sle-condition}
\end{equation}

By some field theoretic calculation as shown the appendix A,
the martingale condition \eqref{mart-SLE} derived in the previous subsection is equivalent to the condition:
\begin{equation}
0=|\chi\ket:=\left(\frac{\kappa}{2}L_{-1}^2-2 L_{-2}+\frac{\tau}{2} \sum_{a=1}^{\dim \g}
J_{\g , -1}^a J_{\g , -1}^a-\frac{\tau (k+h^{\vee})}{2(k+h_a^{\vee})} \sum_{a=1}^{\dim \mathfrak{a}}
J_{\mathfrak{a}, -1}^a J_{\mathfrak{a}, -1}^a\right)|\psi_{\lam_1}\ket.
\label{ns-condition}
\end{equation}
if there exists such bcc operator, where we used the orthogonal basis \eqref{coset-alg}.
To get such bcc operator is a very difficult task, and the only known
example is $SU(2)_k$ and $Z(n)$ parafermion models.

As is the result of the previous subsection, we can construct
the multiple SLEs on the coset WZW models.
So the construction of 1-SLE directly leads to multiple
SLE, but we have to check the construction
by some careful consideration of BCFT or the specific
2d statistical model with boundaries.
We will briefly discuss this point in the next subsection.

Before going to the next section, we make some comment
on another approach to include or exclude the general degrees of freedom.
SLE on $N=1$ minimal models was constructed by Rasmussen \cite{Ras}.
Their formulation is different that of ours in that Brownian motions
on the super manifold directly affect SLE on $H$.
On the other sides, $N=1$ minimal models can be represented as the coset
WZW models by the coset construction.
At this stage, the relation between SLE with Lie group symmetry and that with super symmetry 
is unclear.

%%%%%%%%%%%%%%%%%%%%%%%%%%%%%%%%%%%%%%%%%%%
\subsection{Multiple SLE and $SLE_{\kappa, \rho}$ and BCFT}
%%%%%%%%%%%%%%%%%%%%%%%%%%%%%%%%%%%%%%%%%%%
In this section, we will see the SLE/BCFT correspondence in the multiple SLE
on the coset WZW model.
One of the most important quantity to construct the multiple SLE is the pure
partition function $Z_t$.
$Z_t$ determines the drift terms of multiple SLE.
The explicit form of the drift terms are,
\begin{align}
&dF_\alpha =\kappa dq_\alpha \del_{x_\alpha} \log Z_t+
2\sum_{\substack{\beta=1\\\beta\ne \alpha}}^m
\frac{dq_\beta }{x_\alpha-x_\beta}, \quad
dG_\alpha^a =\frac{\tau}{Z_t}\sum_{\substack{\beta=1\\\beta\ne\alpha}}^m
\frac{s_{\lam_\beta}^a Z_t}{x_\beta-x_\alpha}dq_\alpha.
\end{align}

So we have to calculate $Z_t$. This results in the connection between the multiple SLE
configurations and the operator algebra of BCFT \cite{BBK}.
Before going to the specific construction of $m$-SLE, we mention the conection between the maringale condition
and the KZ equation. The martingale condition \eqref{ns-condition} leads to the KZ equation by multiplying $L_1$. In that sense, KZ equation is the necessary condition of the martingale condition. More explicitly, KZ equation is the weaker equation than the martingale condition in the sense that it cannot determine
the bcc operator of SLE. However it is sufficiently strong in the sense that it can determine the pure partition functions
after the construction of 1-SLE and the determination of bcc operator of  1-SLE.

For the simplest example of the construction of multiple SLE, 2-SLE should be constructed.
For 2-SLE, the appropriate bcc operator at $x_3$ of \eqref{observable} is
strongly limited by the fusion rule of BCFT. In other words, that shows the appropriate
forms of the drift terms should be selected without inconsistency with BCFT.
In this sense, constructing 2-SLE needs the information of the specific BCFT
or the concrete 2d statistical model with boundaries.
After the determination of $\psi_{\lam_3}$, we have to analyze
the 2-SLE by stochastic methods.
This 2-SLE is equivalent to the two famous stochastic processes, the Bessel process
and $SLE_{\kappa, \rho}$. After analyzing them, we have to check the
consistency of the 2-SLE behavior and the BCFT interpretation of the pure
partition function.
If all of these prosedures are accomplished, we can get a well-defined 2-SLE in the
sense it gives BCFT as the distribution function, such as the crossing probability or
the graph connectivity.

The construction of $m$-SLE with $m >2$ requires more conditions.
First, $\psi_{\lam_{m+1}}$ should be determined by fusion rule or ward identity
of BCFT as in the case of 2-SLE. In this sense, checking the consistency of 2-SLE 
and the pure partition function of BCFT is crucial.
After the determination of  $\psi_{\lam_{m+1}}$ corresponding on the specific situation,  
the explicit calculation of the pure partition function should be accomplished.
This calculation may be done by DF integral of BCFT or the combinatorial methods. 
We will discuss the former method in the next section.

All of the present formulation of multiple SLE are applied to a concrete model
i.e. $Z(n)$ parafermion models, in the next section.

%%%%%%%%%%%%%%%%%%%%%%%%%%%%%%%%%%%%%%%%%%%%%%%%%%
\section{$Z(n)$ parafermion model}
%%%%%%%%%%%%%%%%%%%%%%%%%%%%%%%%%%%%%%%%%%%%%%%%%%
 For the explicit example,
we introduce the known results for the $SU(2)_n/U(1)_n$-WZW
or, $Z(n)$ parafermion model.  
The  central charge $c$ 
 and the conformal weight 
$h_{(j,j_z)}$ of the primary operator $\psi_{(j, j_z)}(x)$ in
the isospin-$j/2$ spin-$j_z/2$ representation are, respectively,  written as
\begin{equation}
c=\frac{2(n-1)}{n+2}, \qquad h_{(j, j_z)}=h_{SU(2)_{n},(j.j_z)}-h_{U(1)_{n},j_z }=\frac{j(j+2)}{4(n+2)}-\frac{j_z^2}{4n},
\label{weight}
\end{equation}
{}\eqref{ns-condition} for $Z(n)$ parafermion ($n\ge 4$) case is considered in \cite{Naz1},
and the same condition are derived from different approach in \cite{San}.
By combining their results, the sufficient conditions \eqref{ns-condition} is satisfied 
by the bcc operator which carries isospin-1, spin-0 ($j=2$) \cite{San,PS,PSS}. 
Then, 
\begin{equation}
\kappa=\frac{4(n+1)}{n+3}, \qquad 
\tau=\frac{4n^3}{(n+2)^3} \quad.
\label{kappa}
\end{equation}
This SLE describes the spin boundary interfaces of the model.
The characteristic of this operator is different from the one of minimal and $SU(2)_n$ WZW case.
Fusion rule of this operator has three way branching structure like $\phi_{(2,0)}\phi_{(2,0)}=I+\phi_{(2,0)}+\phi_{(4,0)}$. 
So we have to impose some 
additional conditions because geometric configuration of $m$-SLE does not coincide with the fusion rule. 
In the next subsection we consider this operator as spin boundary operator of $Z(n)$ spin model, and takes 
some physical condition of this situation.
We conjecture this gives correct prediction of pure partition function in this case.
This conjecture may be checked by numerical calculation of corresponding crossing probabilities.

%%
%
%%%%%%%%%%%%%%%%%%%%%%%%%%%%%%%%%%%%%%%%%%%%%%%
\subsection{2-SLEs}
%%%%%%%%%%%%%%%%%%%%%%%%%%%%%%%%%%%%%%%%%%%%%%%
Using the same argument of \cite{BBK, Sak},
we can construct 2-SLE which is equivalent to triangle geometry
with bcc operator on each corner.
After mapping to one corner to $\infty$, the pure partition function is
\begin{equation}
Z=\bra \psi_{\lam_3}(\infty) \psi_\Lambda(x_1) \psi_{\Lambda}(x_2)\ket
=
(x_1-x_2)^\Delta,
\end{equation}
where the exponent is $\Delta=h_{\lam_3}-2h_\Lambda$.   
Setting 
$y_{s}=x_{1t}-x_{2t}$, rescaling the time  by $ds=\kappa dt$,
and inserting it to \eqref{sle-sum} and \eqref{driving-sum}, 
we reduce the processes to the following Bessel process, 

\begin{equation}
dy_s=d B_s+\frac{\Delta+2/\kappa}{y_s}ds.
\end{equation}
The effective dimension of the process is given by $d_{\rm eff}=2 \Delta+4/\kappa+1$. 
This effective dimension determines whether the process is recurrent or not.
The process becomes recurrent when $d_{\rm eff}<2$. It means 2-SLE curves collide on each other.
When $d_{\rm eff}>2$, the process becomes nonrecurrent and 2-SLE curves goes to $\infty$ without
collision.

The fusion rules of spin-1 operators
indicate that $h_{\lam_3}=\frac{6}{n+2}$,  $h_{\lam_3}=\frac{2}{n+2}$ or $h_{\lam_3}=0$ 
. However, $h_{\lam_3}=\frac{2}{n+2}$ is unphysical operator if we identify it as spin cluster boundary.
In terms of SLE that corresponds to the situation that "2-SLE curves fuse to 1-SLE and goes to infinity."
Therefore we exclude $h_{\lam_3}=\frac{2}{n+2}$. (However, there may exist some situations described by this operator.)
Then using the relation \eqref{weight}, one arrives at
$\Delta=2/((n+2))$  or $\Delta=-4/(n+2)$.
Substitution
of the exponent $\Delta$ and $\kappa$ \eqref{kappa} yields
\begin{equation}
d_{\rm eff}=
           2-\frac{3n+2}{(n+1)(n+2)} 
\quad \text{for $h_{\lam_3}=0$}, \quad
d_{\rm eff}=2+\frac{3n+6}{(n+1)(n+2)} \quad
\quad \text{for $h_{\lam_3}=h_{2\Lambda}$}.
\end{equation}
These values are consistent with the BCFT fusion rule, if we recall the previous
analysis of the behavior of the Bessel process.  

For the concluding remark of this section, the former part of the present analysis
of two SLE as Bessel process is valid without detailed analysis of BCFT.
This fact means we can describe spin boundary interface of
$Z(n)$ spin models by both of BCFT and SLE without inconsistency.
However, it does not mean we can straight forwardly get the general boundary interface
of 2d statistical model classified by coset WZW models
because the construction of the drift terms need the BCFT analysis
of the 3 point function and fusion rule of BCFT such as the latter part of our discussion.
Inversely, $SLE_{\kappa ,\rho}$ may describe such boundary curves and the 
parameter $\kappa , \rho$ are determined by level 2 null vector equation,
 conformal dimension and fusion rule of underlying CFT ( with $\rho =\kappa\Delta$).

%%%%%

%%%%%%%%%%%%%%%%%%%%%%%%%%%%%%%%%%%%%%%%%%%%%%%
\subsection{3-SLEs and arch (crossing) probabilities}
%%%%%%%%%%%%%%%%%%%%%%%%%%%%%%%%%%%%%%%%%%%%%%%
Based on the analysis of 2-SLE, we define 3-SLE on $Z(n)$
spin model. Geometrically, 3SLE is equivalent to a rectangular
with bcc operator on each vertex.
The pure partition function, which characterizes the drift terms of the process is
\begin{equation}
Z=\bra \psi_{\lam_4}(x_4)
\psi_{\Lambda}(x_3)\psi_{\Lambda}(x_2)\psi_{\Lambda}(x_1) \ket.
\end{equation}
We take $x_4=\infty$ and the conformal weight $h_{\lam_4}$ is determined by the fusion 
rules. It is 
$h_{\lam_4}=h_{(3,0)}$  or $h_{\lam_4}=h_{(1,0)}$.
Applying the same analysis of 2-SLE to 3-SLE, we excluded
$h_{\lam_4}=h_{(2,0)}, 0$. 
 When the level takes its value in the range $n\ge 4$
and $h_{\lambda_4}=h_{(3,0)}$,
the pure partition function is determined uniquely,
\begin{equation}
Z=\left[(x_2-x_1)(x_3-x_1)(x_3-x_2)\right]^{\frac{2}{n+2}} \quad .
\end{equation}
we expect it to describe the process that 3SLE curves
goes to infinity by previous analysis of  2SLE. 

The case for $h_{\lam_4}=h_{(1,0)}$ needs more detailed analysis
because the pure partition function contains 3 conformal blocks.
However, there only exist two configurations in the rectangular  geometry. 
If we define $[X_1 X_2]$ as the situation when a SLE curve $X_1$ collides with $X_2$,
the possible configurations of 3SLE are $[X_1 X_2][X_3, X_4]$, $[X_1 [X_2 X_3 ]X_4]$. 
Therefore we have to exclude the unphysical conformal block.
The conformal mapping $f(z)=(z-x_1)(x_3-x_4)/((z-x_4)(x_3-x_1))$
simplifies the rectangular geometry, because it 
transforms the points $x_1\to0$,
$x_2\to x$, $x_3\to 1$ and $x_4\to\infty$, where $x=f(x_2)$.
 Thus the
partition function $Z$ can be expressed as
\begin{equation}
Z=((x_2-x_4)(x_1-x_3))^{-2 h_\Lambda}Z(x), \quad
Z(x)=\bra \psi_\Lambda(\infty) \psi_\Lambda(1) \psi_\Lambda(x) \psi_\Lambda(0) \ket.
\label{Z}
\end{equation}
$Z(x)$ should be expressed as the linear combination of the KZ-equation 
\eqref{ns-condition} \cite{KZ}. Assuming two configurations are described by
two conformal blocks,
\begin{equation}
Z(x)=Z_{\rm C_1}(x)+Z_{\rm C_2}(x)
\end{equation}
with left right symmetry,
\begin{equation}
Z_{{\rm C_1}}(x)=Z_{{\rm C_2}}(1-x) .
\label{part}
\end{equation}
The asymptotic behavior of the pure partition function
should be consistent with fusion rule,
\begin{equation}
Z_{{\rm C}_2}(x)\sim \begin{cases}
         x^{h_{2\Lambda}-2h_\Lambda} & \text{ $x\to0$} \\
         (1-x)^{-2h_\Lambda} &\text{ $x\to 1$}
         \end{cases},
\end{equation}
which is consistent with 2-SLE or does not contain unphysical processes.
Under this condition,
 $Z_{\rm C_2}(x)$ is determined by the Dotsenko-Fateev integral,
\begin{align}
Z_{{\rm C_2}}(x)=&x^\frac{2}{n+2}(1-x)^\frac{2}{n+2}\int_1^ \infty dv \int_1^v du
 {v^{-\frac{2}{n+2}}(1-v)^{-\frac{2}{n+2}}(v-x)^{-\frac{2}{n+2}}} \nn \\
&{u^{-\frac{2}{n+2}}(1-u)^{-\frac{2}{n+2}}(u-x)^{-\frac{2}{n+2}}(v-u)^{\frac{2}{n+2}}
 f(u,v)
} 
\end{align}
with
\begin{equation}
f(u,v)=\left(\frac{1}{(v-x)(u-1)}+\frac{1}{(v-1)(u-x)}+\frac{1}{v(u-1)}+\frac{1}{(v-1)u}+\frac{1}{v(u-x)}+\frac{1}{(v-x)u}\right) .
\end{equation}
The significant feature of the Dotsenko-Fateev integral we use is the representation of conformal block
by the contour of integral.
Unfortunately this function contains divergence at $z=1$, and hence analytical continuation
is necessary. 

This pure partition function can give the explicit form of the crossing probability \cite{BBK,Koz}
which describes the realization probability of each configuration,
\begin{equation}
\mathbf{P}[{{\rm C}_i}]=\frac{Z_{{\rm C}_i}(x)}{Z_{{\rm C}_1}(x)+Z_{{\rm C}_2}(x)}, \quad
i=1, 2.
\end{equation}
In other words, the validity of our construction of the pure partition function can
be checked by the calculation of this crossing probability by other methods such as
Monte-Carlo simulation and combinatorics.
We will discuss this point in the next section.

%
%
%%%%%%%%%%%%%%%%%%%%%%%%%%%%%%%%%%%%%%%%%%%%%%%%%%%%%%%%%%%%%%%%%%%
\section{Conclusion}
%
%%%%%%%%%%%%%%%%%%%%%%%%%%%%%%%%%%%%%%%%%%%%%%%%%%%%%%%%%%%%%%%%%

Based on conformal field theory, we formulated the multiple SLEs on the coset WZW models, and determined the
 drift terms which satisfy martingale condition. This fact infers BCFT or 2d statistical model with multiple domains wall can be constructed
 by the distribution
functions of the multiple SLEs under the existence of  the single SLEs. It is an evidence of SLE/CFT correspondence in
the general WZW models.   

Especially, we constructed 2-SLE on $Z(n)$ spin model whose pure partition function is consistent with that of SLE. Based on the construction of 2-SLE, we calculated the crossing
probability by the calculation of  the 4-point function which verifies the construction of the 3-SLE. This formal 
integral contains divegence at $x=1$, but we believe the analytical continuation may make sense
(as in usual hyper geometric function). The author is working with that problems, and the forth coming paper
will include analytical and numerical result \cite{AFPS}.

The drift terms of $Z(n)$ spin model contain the pure partition function of spin-$1$ fields.
Therefore we have to choose the physical process in the fusion diagram. That is a different point from single SLE. We can extract the
physical or SLE realizable processes in general $SU(2)_{k}$ coset WZW model if the single and two SLE are formulated.
In that sense, analysis of 2-SLE is fundamental and related to well definedness of one point function with
the bcc operator. Moreover 2-SLE is the key ingredient of the construction of conformal loop ensemble.
However in more general WZW models, such as $SU(3)_{k}$ WZW model, such formulation may be difficult because
of the complicated fusion rules. This fact may be related to the impossibility of the construction of 1 SLE in $SU(N)_k$
with $N$ bigger than $2$ for general $k$, but the fundamental meaning is unclear. 

In appendix, we have shown Lie group symmetry may change the probability measures of the multiple SLE.
This fact indicates SLE does not determined only by local fractal dimension of the curve\cite{MSW}. 

Finally, there are a lot of open problems related to SLE and WZW models.
We do not know the construction of SLE curves in more general statistical mechanical models, such as coupled Potts, and network models.
At lattice level we naively expect the boundary inteface of such models may be described by SLE,
but no analysis of such fractal objects. Moreover the null vector condition of the b.c.c. operator to describe SLE seems to be
very complicated in such cases.
Our formulation may be helpful in such case because our formulation ensures the probability measures such as the crossing probability are described by BCFT correlation functions. Such information may be helpful to guess boundary condition changing operator which describes SLE curve numerically and analytically.

%%%%%%%%%%%%%%%%%%%%%%%%%%%%%%%%%%%%%%%%%%%%%%%%%%%%%%%%%%%%%%%%
%%%%%%%%%%%%%%%%%%%%%%%%%%%%%%%%%%%%%%%%%%%%%%%%%%
\section*{Acknowledgment}
%%%%%%%%%%%%%%%%%%%%%%%%%%%%%%%%%%%%%%%%%%%%%%%%%%%%%%%%%%%%%%%%
The author really appreciates Kazumitsu Sakai for fruitful discussions
and critical reading of the manuscript.
The author also thanks Raoul Santachiara for a lot of helpful discussions 
at  {\it quantum integrable systems, 
conformal field theories, Stochastic processes} (12-23 september 2016) 
held in Carg\`ese, where a part of this
work was presented. The author also thanks Masaki Oshikawa
for discussion and encouragement at the first period of this work.
%
%%%%%%%%%%%%%%%%%%%%%%%%%%%%%%%%%%%%%%%%%%%%%%%%%
\appendix
\def\thesection{\Alph{section}}
\def\reference{\relax\refpar}

\section{Necessary information of coset WZW models}
First, we define algebraic relations of Virasoro algebra and level $k$ $G/A$ Kac-Moody algebra \cite{KZ, KSS},
\begin{align}
&\left[J^{a}_{\g, n}, J^b_{\g, m}\right]=\sum_c i f^{ab}_{\g\,\,\,\,\,c} J^c_{\g,n+m}+k n \delta^{a,b}\delta_{n+m,0},
                 \nn  \\
&\left[J^a_{\mathfrak{a},n}, J^b_{\mathfrak{a},m}\right]=\sum_c i f^{ab}_{\mathfrak{a}\,\,\,\,\,c} J^c_{\mathfrak{a},n+m}+k n \delta^{a,b}\delta_{n+m,0},
                 \nn  \\
&\left[L_n,J_{\mathfrak{a},m}^a\right]=0, \nn \\
& \left[L_n,L_m\right]=(n-m)L_{n+m}+\frac{c}{12}(n^3-n)\delta_{n+m,0}.
\label{coset-alg}
\end{align}
In this section, we use the orthogonal basis. The central charge is given by $c=\frac{k \dim \g}{k+h^{\vee}_{\g}}-\frac{k \dim \mathfrak{a}}{k+h^{\vee}_{\mathfrak{a}}}$.

Then we assume the Sugawara construction,  
\begin{equation}
T(z)=\frac{1}{2(k+h^{\vee}_{\g})}\sum_{a=1}^{\dim \g}
:J^{a}_{\g}(z) J^{a}_{\g}(z):-\frac{1}{2(k+h^{\vee}_{\mathfrak{a}})}\sum_{a=1}^{\dim \mathfrak{a}}
:J^{b}_{\mathfrak{a}}(z) J^b_{\mathfrak{a}}(z):
\end{equation}
For simplicity, we assume following OPE,
\begin{align}
&T(z)\phi_{(\lambda,\nu)}(w)\sim\frac{h_{(\lambda,\nu )} \phi_\lambda(w)}{(z-w)^2}+
    \frac{\del_w \phi_\lambda(w)}{z-w}, \quad \nn \\
&J_{\g}^a(z) \phi_{(\lambda,\nu)}(w)\sim\frac{-t_{\g, \lambda}^a \phi_{(\lambda,\nu)}(w)}{z-w}, \nn \\
&J_{\mathfrak{a}}^b(z) \phi_{(\lambda,\nu)}(w)\sim\frac{-{t}_{\mathfrak{a}, \nu}^b \phi_{(\lambda,\nu)}(w)}{z-w},
\end{align}
The conformal dimension is
\begin{equation}
h_{(\lambda,\nu )}=h_{\g,\lambda} -h_{\mathfrak{a},\nu}=\frac{(\lam,\lam+2\rho_{\g})}{2(k+h^{\vee}_{\g})}-\frac{(\nu,\nu+2\rho_{\mathfrak{a}})}{2(k+h^{\vee}_{\mathfrak{a}})}.
\end{equation} 
$\rho_{g}$, $\rho_{\mathfrak{a}}$ are Weyl vector.
Substituting the difinition of $L$ and $J$, we get the following form of differential equation of the operator $X(\{ z_i \} )=\prod_{i}^{m}\phi_{(\lambda_{i},\nu_{i})}(z_i)$ :
 \begin{alignat}{2}
&\bra (L_{-l}\psi_\Lambda)(z) X\ket=\mathcal{L}_{-l} \bra \psi_\Lambda(z) X \ket,
\quad &&\mathcal{L}_{-l}=\sum_{j=1}^m \left[\frac{(l-1)h_{(\lam_j,\nu)}}{(z_j-z)^l}
-\frac{\del_{z_j}}{(z_j-z)^{l-1}}\right], \nn \\
&\bra (J_{\g, -l}^a \psi_\Lambda)(z) X\ket=\mathcal{J}_{\g, -l}^a\bra \psi_\Lambda(z) X \ket,
\quad && \mathcal{J}_{-l}^a=\sum_{j=1}^m \frac{t_{\g, \lam_j}^a}{(z_j-z)^l}, \nn \\
&\bra (J_{\mathfrak{a},-l}^b \psi_\Lambda)(z) X\ket=\mathcal{J}_{\mathfrak{a},-l}^b\bra \psi_\lam(z) X \ket,
\quad && \mathcal{J}_{\mathfrak{a}, -l}^b=\sum_{j=1}^m \frac{t_{\mathfrak{a}, \nu_j}^b}{(z_j-z)^l}.
\label{LJ-dif}
\end{alignat}
In this paper, we assumed OPE, but there is other formulations assuming the commutation relation or Ward identity.

Finally, the KZ equation which determines the correlation functions is,
\begin{equation}
\left(L_{-1}+\frac{1}{k+h_{\g}^{\vee}}\sum_{a=1}^{\dim \g} t_{\g,\lam}^a 
J_{\g, -1}^a-\frac{1}{k+h_{\mathfrak{a}}^{\vee}}\sum_{b=1}^{\dim \mathfrak{a}} t_{\mathfrak{a},\nu}^b 
J_{\mathfrak{a}, -1}^b\right) |\phi_{(\lam, \nu)}\ket=0,
\label{coset-KZ}
\end{equation}
  
\section{Wakimoto free field representation}
In this part, we briefly review the Wakimoto free field representation of
$SU(2)_k$ WZW model by Dotsenko \cite{Dot}.

It contains three bosonic fields,
($\omega$, $\omega^{+}$, $\phi$), and "back ground charge" $\alpha_0$,
with $k=-2-\frac{1}{2\alpha_{0}^2}$.
\begin{equation}
\bra \omega (z) \omega ^+ (z') \ket=-\bra \omega^+ (z) \omega  (z')\ket=\frac{i}{z-z'}.
\bra \phi (z) \phi  (z')\ket=\log \frac{1}{z-z'}
\end{equation}
$SU(2)_k$ currents are constructed by following expression.
\begin{equation}
J^{+} (z)=\omega^{+} (z).
\end{equation}
\begin{equation}
J^{0} (z)=-i(\omega \omega^{+}+\frac{1}{2\alpha_0}\phi ').
\end{equation}
\begin{equation}
J^{-} (z)=\omega\omega\omega^{+} +ik\omega ' +\frac{1}{\alpha_0} \phi ' \omega.
\end{equation}
Moreover primary fields are defined by,
\begin{equation}
\phi^{j}_{m}(z)=(\omega (z))^{j-m} e^{-2ij\alpha_0 \phi (z)}.
\end{equation}
The screening operator is the formal integral of $\omega^{+}(z) e^{2i\alpha_0 \phi (z)} $.

Then by Sugawara construction and Ward identity, we can obatain multipoint
functions.
The charge condition of the correlation function is,
\begin{equation}
N_{+}-N_{-}=s, \sum{\alpha_i}=2\alpha_{0}s, s=-k-1
\end{equation}
Where $N_{+}$ and $N_{-}$ represent the total number of $\omega^{+}$, $\omega$ and
$\alpha_i$ is the charge of the vertex operator.

We note here the necessary part of the conjugate field which we use in the section 3: 
\begin{equation}
\widetilde{\phi^{1}_{0}} \sim (\omega^{+})^{s+2} (\omega) e^{2i\alpha_{0}(s+1)\phi}
\end{equation}

\section{Correspondence of the crossing probability of $SU(2)_2$ WZW model and Ising model }

The martingale boundary operator of $SU(2)_k$ WZW model is $\phi^1_{\pm 1}$
$\kappa=\frac{4(k+2)}{k+3}$, $\tau=\frac{2}{k+3}$ \cite{ABI}.
The pure partition function which gives the crossing probability is given by
\begin{equation}
Z_{{\rm C_2}}(x)=x^\frac{1}{2(k+2)}(1-x)^\frac{1}{2(k+2)}F(\frac{k+3}{k+2},\frac{3}{k+2};\frac{k+4}{k+2}:x),
\end{equation}
and left right symmetry condition.

Especially, we think about $SU(2)_2$ WZW model. Then $\kappa=\frac{16}{5}$, $\tau=\frac{2}{5}$.
Hence the pure partition function is given by
\begin{equation}
Z_{{\rm C_2}}(x)=x^\frac{1}{8}(1-x)^\frac{1}{8}F(\frac{5}{4},\frac{3}{4};\frac{3}{2}:x),
\end{equation}
and left right symmetry. This model may describe some boundary of RSOS model of Fendley's model\cite{JMO, Fen}.

As is well known, the same crossing probability is obtained from Ising model. That is from the consequence of
coset construction of the model. That means, the local characteristics of $SU(2)_2$ WZW model
 ($\kappa=\frac{16}{5}$, $\tau=\frac{2}{5}$.)  are different from that of Ising model ($\kappa=\frac{16}{3}$, $\tau=0$.),
 but the global characteristics are the same.   

What we observed in this paper is the role of Lie group symmetry in SLE. 1SLE processes are not affected by
additional Lie group symmetry at geometric level, but multiple or SLE with addtional variables are changed by
this symmetry if the consistency with BCFT is required. In that sense, $SLE_{\kappa,\tau}$
may describe boundary interfaces of 2d statistical model with additional symmetry.  

\section{$Z(n)$ spin model}
In this paper, we considered the Fateev-Zamolodchikov(FZ) point of $Z(n)$ spin model \cite{ZF}.
First we think about square lattice and assume spin $\sigma_i$ lives in  each sites $i$.
Each spin takes $N$ values from $0$ to $n-1$.

Partition function of $Z(n)$ spin model is given by
\begin{equation}
Z=\sum_{\sigma_i  } {\rm{exp}}\left( -\beta\sum_{\bra i,j\ket} H\left(\sigma_i -\sigma_j \right) \right), H(\sigma)=-\sum_{m=1}^{\lfloor \frac{n}{2} \rfloor}
J_{m} ({\rm{cos}} (\frac{2\pi m\sigma}{n})-1).
\end{equation}

And the FZ point is described by
\begin{equation}
x_0=1, x_n=\prod_{k=1}^{n-1}{\frac{{\rm{sin}}(\frac{\pi k}{N}+\frac{\pi}{4N})}{{\rm{sin}}(\frac{\pi(k+1)}{N}-\frac{\pi}{4N})}}
\end{equation}
with $x_n={\rm{exp}} (-\beta H(n))$.

%%%%%%%%%%%%%%%%%%%%%%%%%%%%%%%%%%%%%%%%%%%

\end{document}